\documentclass[letterpaper,10pt]{article} 

\usepackage{opticameet3} %
\usepackage{bm}
\usepackage{mleftright}
\newcommand{\mlr}[1]{\mleft(#1\mright)}

\usepackage[nohyperlinks,nolist]{acronym}
\acrodef{AWGN}[AWGN]{additive white Gaussian noise}

\acrodef{B2B}[B2B]{back-to-back}
\acrodef{BER}[BER]{bit error rate}
\acrodef{BMI}[BMI]{binary mutual information}
\acrodef{BPS}[BPS]{blind phase search}

\acrodef{CD}[CD]{chromatic dispersion}
\acrodef{CM}[CM]{constant-modulus}
\acrodef{CMA}[CMA]{constant-modulus algorithm}
\acrodef{CNN}[CNN]{convolutional neural network}
\acrodef{CPE}[CPE]{carrier-phase estimation}

\acrodef{DD}[DD]{decision-directed}
\acrodef{DFE}[DFE]{decision feedback equalizer}
\acrodef{DP}[DP]{dual-polarization}
\acrodef{DSP}[DSP]{digital signal processing}
\acrodef{DFB}[DFB]{distributed-feedback laser}

\acrodef{EAM}[EAM]{electro-absorption modulator}
\acrodef{ELU}[ELU]{exponential linear unit}
\acrodef{ELBO}[ELBO]{evidence lower bound}

\acrodef{FEC}[FEC]{forward error correction}
\acrodef{FIR}[FIR]{finite impulse response}
\acrodef{FPGA}[FPGA]{field-programmable gate array}
\acrodef{FTTH}[FTTH]{fiber-to-the-home}

\acrodef{GRU}[GRU]{gated recurrent unit}

\acrodef{IM/DD}[IM/DD]{intensity-modulation and direct-detection}
\acrodef{IP}[IR]{impulse response}
\acrodef{ISI}[ISI]{inter-symbol interference}

\acrodef{KL}[KL]{Kullback-Leibler}

\acrodef{LDPC}[LDPC]{low-density parity-check}
\acrodef{LLR}[LLR]{log-likelihood ratio}
\acrodef{LMS}[LMS]{least mean squares}

\acrodef{MA}[MA]{moving average}
\acrodef{MAP}[MAP]{maximum a~posteriori}
\acrodef{MIMO}[MIMO]{multiple-input multiple-output}
\acrodef{ML}[ML]{maximum likelihood}
\acrodef{MMA}[MMA]{multi-modulus algorithm}
\acrodef{MMSE}[MMSE]{minimum mean squared error}
\acrodef{MSE}[MSE]{mean squared error}

\acrodef{NN}[NN]{neural network}
\acrodef{NRZ}[NRZ]{non-return-to-zero}

\acrodef{OOK}[OOK]{on-off-keying}
\acrodef{ONU}[ONU]{optical network unit}
\acrodef{OLT}[OLT]{optical line terminal}

\acrodef{PAM4}[PAM4]{4-ary pulse amplitude modulation}
\acrodef{PCS}[PCS]{probabilistic constellation shaping}
\acrodef{pdf}[pdf]{probability density function}
\acrodef{PMD}[PMD]{polarization mode dispersion}
\acrodef{pmf}[pmf]{probability mass function}
\acrodef{PON}[PON]{passive optical network}
\acrodef{PSK}[PSK]{phase shift keying}
\acrodef{PSP}[PSP]{principal state of polarization}

\acrodef{QAM}[QAM]{quadrature amplitude modulation}
\acrodef{QoS}[QoS]{quality of service}

\acrodef{ReLU}[ReLU]{rectified linear unit}
\acrodef{RDE}[RDE]{radius-directed equalizer}
\acrodef{ROP}[ROP]{received optical power}
\acrodef{RRC}[RRC]{root-raised cosine}
\acrodef{rvm}[rvm]{real-valued multiplication}

\acrodef{SER}[SER]{symbol error rate}
\acrodef{SSMF}[SSMF]{standard single mode fiber}
\acrodef{SNR}[SNR]{signal-to-noise ratio}
\acrodef{SOA}[SOA]{semiconductor optical amplifiers}
\acrodef{sps}[sps]{samples per symbol}

\acrodef{VAE}[VAE]{variational autoencoder}
\acrodef{VOA}[VOA]{variable optical attenuator}
\acrodef{VQVAE}[VQVAE]{vector-quantized variational autoencoder}

\usepackage{float}
\usepackage{wrapfig}
\restylefloat{figure}
\usepackage[font=footnotesize]{caption}
\usepackage{subcaption}
\usepackage{comment}

\newcommand\authormark[1]{\textsuperscript{#1}}

\usepackage[colorlinks=true,bookmarks=false,citecolor=blue,urlcolor=blue]{hyperref} %
\usepackage{siunitx}
\begin{document}

\title{Fully-blind Neural Network Based Equalization for Severe Nonlinear Distortions in 112 Gbit/s Passive Optical Networks}

\author{Vincent Lauinger,\authormark{1,*} Patrick Matalla,\authormark{2} Jonas Ney,\authormark{3} Norbert Wehn,\authormark{3} Sebastian Randel,\authormark{2} and Laurent Schmalen\authormark{1}}

\address{\authormark{1}Communications Engineering Lab (CEL), Karlsruhe Institute of Technology (KIT), 76131 Karlsruhe, Germany\\ \authormark{2}Institute of Photonics and Quantum Electronics (IPQ), Karlsruhe Institute of Technology (KIT), 76131 Karlsruhe, Germany\\
\authormark{3}Microelectronic Systems Design (EMS), RPTU Kaiserslautern-Landau, 67653 Kaiserslautern, Germany} %

\email{\authormark{*}\texttt{vincent.lauinger@kit.edu}} %

\begin{abstract}
We demonstrate and evaluate a fully-blind \ac*{DSP} chain for 100G~\acp*{PON}, and analyze different equalizer topologies based on neural networks with low hardware complexity.
\end{abstract}

\vspace*{-0.0ex}
\section{Introduction}\vspace*{-1.5ex}
In recent decades, \acp{PON} have been the key technology to enable broadband internet access in cities worldwide. Since \acp{PON} are primarily used for \ac{FTTH}, the end-user transceivers must be cheap and power efficient while covering the increasing demand of %
data rates. For this reasons, they typically rely on \ac{IM/DD} of the optical signal. Current research is focusing on data rates beyond recent 50G-\ac{PON} standardization efforts~\cite{bonk50G}, i.e., towards \acp{PON} which are capable of delivering \SI{100}{Gbit/s}~\cite{bonk100G}. Since cost-effective hardware hinders increasing the symbol rate, the focus shifts towards higher-order modulation formats such as \ac{PAM4}. %
However, compared to conventional \ac{OOK}, which is used until 50G-\ac{PON}, multi-level modulation formats are more prone to nonlinerities and, due to its reduced \ac{SNR} tolerance, require optical amplification. 
The utilized low-cost \ac{SOA} distort the signal at high \ac{ROP} due to nonlinear gain saturation, which reduces the dynamic range~\cite{murphy}. Additionally, \acp{EAM} distort the signal at the transmitter due to their nonlinear electro-optical power transfer function and, %
\ac{CD} corresponds to a nonlinear effect in an \ac{IM/DD} link, which %
limits high-speed \acp{PON}, even in the O-band~\cite{bonk50G}.

Digital linear equalization, as introduced in the 50G-\ac{PON} standard, can compensate for such nonlinear distortions only to a certain extent. Thus, nonlinear equalization methods draw attention of the community. In particular, machine learning using \acp{NN}, which proved to be highly capable of compensating for nonlinear impairments, is a contender for future equalizers. 
Current work focuses %
on the correction of isolated effects such as distortions by the \ac{SOA}~\cite{murphy}, or \ac{CD}~\cite{houtsma, yi}. %
However, all of these works considered supervised learning meaning that they require known training sequences or intense offline learning. Hence, these algorithms cannot adapt to varying transmission parameters or require pilot sequences, which significantly reduce the net data rate. 

This work contains two major contributions. Firstly, we apply a novel blind adaptive learning algorithm inspired by a \ac{VQVAE}~\cite{songTCOMM} and combine it with a blind \ac{DSP} chain~\cite{Matalla}, which allows us to transmit fully-blindly \SI{112}{Gbit/s} through a %
\ac{PON} upstream with an \ac{EAM} at the \ac{ONU} and with an \ac{SOA} at the \ac{OLT}. %
Precisely, we transmit \ac{PAM4} at \SI{56}{GBd} for a distance of \SI{2.2}{km} in C-band ($D$~$\approx$~$\SI{15.5}{ps/nm/km}$), which corresponds to approx. \SI{9}{km} in O-band upstream at \SI{1270}{nm} ($D$~$\approx$~$\SI{-3.7}{ps/nm/km}$),
and investigate the equalization algorithm under all impairments mentioned above. 
Secondly, we investigate \ac{NN} topologies which are reasonable from a hardware implementation perspective, and compare them to other state-of-the-art \ac{NN} topologies such as \acp{GRU}.

\vspace*{-1ex}
\section{Blind Adaptive \ac{NN}-based Equalizers} \vspace*{-1.5ex}

A promising algorithm to blindly and adaptively update \ac{NN}-based equalizers is the recently-proposed \ac{VQVAE}-inspired learning algorithm~\cite{songTCOMM}.
While it can be derived from statistics and the concept of variational inference, which enables \ac{ML} approximating estimation, it boils down to a rather simple loss function used to update the equalizer parameters,  $\mathcal{L} = \beta \ ||\tilde{\bm{x}} - \hat{\bm{x}}||^2 + (1-\beta) \ ||\bm{y} - f_\theta\mlr{\hat{\bm{x}}}||^2 $ ,
where $\beta$ is a weighting factor, $\bm{y}$ are the received samples, $\tilde{\bm{x}}$ is the equalizer output, $\hat{\bm{x}}$ are the expected symbols after hard decision, and $f_\theta\mlr{\cdot}$ is a channel estimator, e.g., another \ac{NN}, which is also learned during training. %
As a byproduct, latter provides valuable information for, e.g., \ac{QoS} estimation. 
In fact, the loss function consists of two terms, where the first (commitment loss) comes down to a \ac{DD} \ac{LMS} loss while the second term (reconstruction loss) tries to reconstruct the received signal from the detected signal using the learned channel estimate. While the latter is especially important for the startup phase, the weighting factor $\beta$ allows to shift more towards the \ac{DD}-\ac{LMS} loss after convergence. For this reason, we employ a scheduler which increases $\beta$ by a factor of $a_\beta$ after every $N_\beta$-th training iteration. We use Adam~\cite{kingma2014adam}, a gradient descent based optimizer, and another scheduler with $a_{\mathrm{lr}}$ and $N_{\mathrm{lr}}$ to control the learning rate. Per sequence, we have 7600 training iterations of 720 samples each at two \ac{sps} per iteration.
All remaining hyperparameters have been optimized at $\mathrm{ROP} = \SI{-2}{dBm}$ %
using a Bayesian search. 

\begin{figure}[t]
    \centering
    \includegraphics{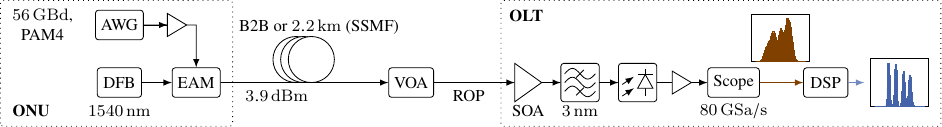}	
    \caption{\vspace*{-4.5ex}Experimental setup for the \SI{112}{Gbit/s} (\SI{56}{GBd}) PON upstream through a \ac*{SSMF} in the C-band.} %
    \label{fig:experimental_setup}
    \vspace*{-3ex}
\end{figure}

\begin{figure} [!b]
		\centering
        \vspace*{-3ex}
		\includegraphics{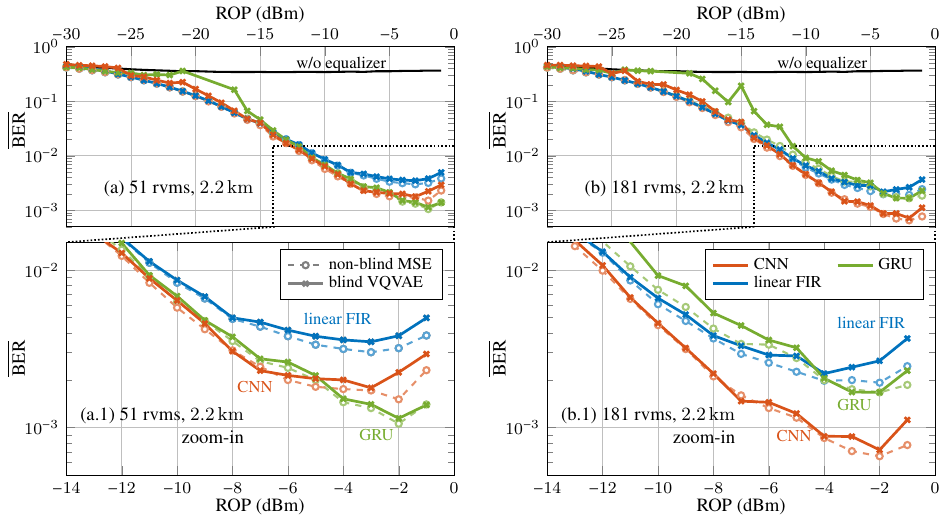}
		\vspace*{-1ex}
		\caption{Median $\overline{\text{BER}}$ for a fiber length of \SI{2.2}{km} (C-band) with \acp{NN} of 51~\acp*{rvm}~(a) and 181~rvms~(b) .}
		\label{Results:BER_ROP_matrix}
		\vspace*{-4ex}
\end{figure} %

We consider two main types of \acp{NN}, namely \acp{GRU} and \acp{CNN}, which are compared to a commonly-used linear \ac{FIR} filter. The \ac{NN} sizes are chosen to be still feasible for practical implementation.
For the \ac{GRU}, we took the suggested topology from~\cite{murphy}, where it was also analyzed for \acp{PON}. Precisely, we chose a large \ac{GRU} with 180~\acp{rvm} per symbol (3 input taps, 6 hidden \ac{GRU} cells and one output tap), and a small with 51~\acp{rvm} (8 input taps, 6 hidden \ac{GRU} cells and 6 output taps). 
Note that latter has more learnable parameters, but reduces the \acp{rvm} by moving the input block by six samples (instead of one sample) each time and outputting six equalized symbols at once. 
To accommodate for the oversampled input, the output sequence of the \ac{GRU} is downsampled by $\text{sps}=2$. 
It also should be mentioned that \acp{GRU} have in general an overhead in the required hardware resources due to the necessity of storing the state, the need of a hyperbolic tangent (tanh) and two sigmoid activation functions and the recurrent paths which constrain parallelization and introduce a routing overhead~\cite{RNN_HW}. In contrast, the \acp{CNN} are easily implementable on a \ac{FPGA}~\cite{neySAMOS} and consist of three one-dimensional layers with simple \ac{ReLU} activation (after the first and second layer). The \acp{CNN} reduce the required \acp{rvm} by strided convolutions with $\text{stride}= (2,2,2+\text{sps})$ and output 8 equalized samples simultaneously. In fact, we choose a small \ac{CNN} of 49~\acp{rvm} with a kernel size $K=(7,5,7)$ and $C=(4,4,8)$ output channels per layer in a way that it allows implementation on the high-performance FPGA Xilinx XCVU13-P with the required throughput of \SI{56}{GBd}~\cite{neySAMOS}. For comparison, we also implemented a large \ac{CNN} of 180~\acp{rvm} with a kernel size $K=(11,9,9)$ and a number of output channels per layer $C=(7,7,8)$, as well as \ac{FIR} filters of similar complexity with 51~taps and 181~taps. 

The channel estimator of the \ac{VQVAE} equalizer is implemented as two-layer \ac{CNN} with $K=(11,15)$, and $C=(5,1)$, which is large enough to ensure proper channel estimation capabilities.

\vspace*{-1ex}
\section{Experimental Setup and Results}\vspace*{-1.5ex}
In Fig.~\ref{fig:experimental_setup}, we display the experimental setup, which emulates a \ac{PON} upstream path. At the \ac{ONU} side, we use a Keysight USPA DAC3 to generate a \SI{56}{Gbaud} \ac{NRZ} \ac{PAM4} signal with a sequence length of $2^{19}$ symbols and amplify it to a peak-to-peak voltage of \SI{2}{V} to achieve a sufficient optical modulation amplitude at the low-cost \ac{EAM}, which outputs an optical transmit power of \SI{3.9}{dBm}.
A \ac{DFB} provides the optical carrier at a wavelength of \SI{1540}{nm}. %
\begin{wrapfigure}[12]{r}{0.48\textwidth}
	\vspace*{-0.7ex}
	\raggedright
	\includegraphics{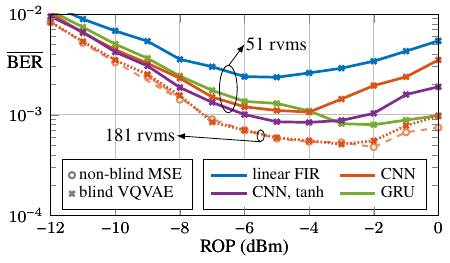} 
	\vspace*{-5ex}
	\caption{Median $\overline{\text{BER}}$ for optical \ac{B2B}.} \label{Results:BER_ROP_0km_10km}
\end{wrapfigure}
While propagating through the fiber, the signal accumulates \ac{CD} and is attenuated by a \ac{VOA} to set a certain \ac{ROP}. At the receiver on the \ac{OLT}-side, an \ac{SOA} with a \SI{3}{nm}-bandpass filter accounts for the higher \ac{SNR} requirements of \ac{PAM4} compared to \ac{OOK}. 
We use a \SI{40}{GHz}-photodiode with conventional amplifier since there was no avalanche photodiode with transimpedance amplifier available in our lab during the time of the experiments, which would further improve the receiver sensitivity. Finally, the electrical signal is captured by a \SI{33}{GHz}-real-time oscilloscope at \SI{80}{GSa/s} and resampled to two \ac{sps} for the subsequent blind feedforward clock recovery suitable for \acp{PON}~\cite{Matalla} and the equalization. %
Either the blind \ac{VQVAE} (indicated by cross markers) learning algorithm or, as a reference, the standard supervised (non-blind) \ac{MSE} loss (indicated by circle markers) is used to update the equalizer taps. The colors indicate different equalizer topologies.
After every one-hundredth training iteration, the \ac{BER} is estimated over a test set of \num{50000}~symbols (cut from the same recorded sequence) and the last 10 estimates are averaged to get the mean $\overline{\text{BER}}_\mathrm{Seq}$ per sequence. 
In total, 15 sequences are recorded per working point, and the median $\overline{\text{BER}}$ is displayed in Fig.~\ref{Results:BER_ROP_matrix}. %

For the same number of \acp{rvm}, the \ac{NN} based equalizers outperform the linear \ac{FIR} filters. %
In the strongly non-linear regime of $\text{ROP}>\SI{-3}{dBm}$, the \ac{GRU} with 51~\acp{rvm} %
reaches the lowest $\overline{\text{BER}}$, while the \ac{CNN}, which has a less complex ReLU activation function, has advantage in the more linear regime. 
\begin{wrapfigure}[12]{R}{0.48\textwidth}
	\vspace*{-4ex}
	\raggedright
	\includegraphics{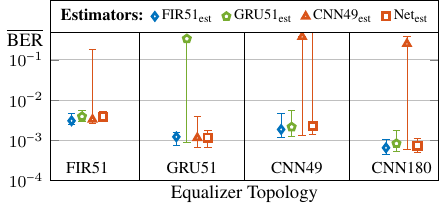} 
	\vspace*{-2.5ex}
	\caption{Median $\overline{\text{BER}}$ (markers) and the corresponding error bars (to the best and worst estimates per 15 captured sequences) at \SI{2.2}{km}, \SI{-2}{dBm} for the proposed equalizers, trained by the blind VQVAE with different channel estimator nets.} \label{Results:BER_Est}
\end{wrapfigure}
Overall, the \ac{CNN} with 180~\acp{rvm} performs the best. For all topologies, the blind \ac{VQVAE} algorithm approaches the performance of the non-blind \ac{MSE} loss. %
Similar results can be seen in Fig.~\ref{Results:BER_ROP_0km_10km} for optical \ac{B2B}. %
For the \ac{CNN} with 49~\acp{rvm}, the usage of a tanh activation function (purple) increases the performance.  

Different channel estimator topologies for the \ac{VQVAE} algorithm at \SI{2.2}{km} and \SI{-2}{dBm} are analyzed in Fig.~\ref{Results:BER_Est} for the proposed equalizer topologies. The markers indicate the different estimators, which topologies are based on the similarly-named equalizers. %
\textit{NetEst} corresponds to the above-described two-layered estimator \ac{CNN}, which has been used so far along with the \ac{VQVAE}. %
Different estimators work well for different equalizers, while the FIR based estimator with only 51~\acp{rvm} performs good in general. This indicates that the estimator can be significantly simplified and optimized.%

\vspace*{-1ex}
\section{Conclusion}\vspace*{-1.5ex}
We demonstrate the fully-blind transmission of \SI{112}{Gbit/s} through a \ac{PON} uplink path with \ac{NN} topologies, which complexity is low enough for implemention on state-of-the-art \acp{FPGA}. Furthermore, we show that appropriate \ac{NN} topologies reach a lower \ac{BER} as classical linear \ac{FIR} filters of similar computational complexity, and that the novel blind and adaptive \ac{VQVAE}-based learning algorithm is capable of approaching the performance of the non-blind \ac{MSE} loss.

\vspace*{2ex}
\footnotesize 
\noindent{\bf{Acknowledgements}}:
This work was carried out in the framework of the CELTIC-NEXT project AI-NET-ANTILLAS (C2019/3-3, grant 16KIS1316 and 16KIS1317) and the project KIGLIS (grant 16KIS1228), both funded by the German Federal Ministry of Education and Research (BMBF). The authors thank Jinxiang Song from Chalmers University of Technology for valuable discussions.

\end{document}